# FPGA Implementation of Convolutional Neural Network for Real-Time Handwriting Recognition

Shichen (Justin) Qiao, Haining Qiu,
Lingkai (Harry) Zhao, Qikun Liu, Eric J. Hoffman

{sqiao6, hqiu37, lzhao224, qliu292, erichoffman}@wisc.edu

# Abstract

Machine Learning (ML) has recently been a skyrocketing field in Computer Science. As computer hardware engineers, we are enthusiastic about hardware implementations of popular software ML architectures to optimize their performance, reliability, and resource usage. In this project, we designed a highly-configurable, real-time device for recognizing handwritten letters and digits using an Altera DE1 FPGA Kit. We followed various engineering standards, including IEEE-754 32-bit Floating-Point Standard, Video Graphics Array (VGA) display protocol, Universal Asynchronous Receiver-Transmitter (UART) protocol, and Inter-Integrated Circuit (I2C) protocols to achieve the project goals. These significantly improved our design in compatibility, reusability, and simplicity in verifications. Following these standards, we designed a 32-bit floating-point (FP) instruction set architecture (ISA). We developed a 5-stage RISC processor in System Verilog to manage image processing, matrix multiplications, ML classifications, and user interfaces. Three different ML architectures were implemented and evaluated on our design: Linear Classification (LC), a 784-64-10 fully connected neural network (NN), and a LeNet-like Convolutional Neural Network (CNN) with ReLU activation layers and 36 classes (10 for the digits and 26 for the case-insensitive letters). The training processes were done in Python scripts, and the resulting kernels and weights were stored in hex files and loaded into the FPGA's SRAM units. Convolution, pooling, data management, and various other ML features were guided by firmware in our custom assembly language. This paper documents the high-level design block diagrams, interfaces between each System Verilog module, implementation details of our software and firmware components, and further discussions on potential impacts.

## 1. Important Notes

Hardware Usage:
- (1) Altera DE1 Development and Education Board:
    https://www.terasic.com.tw/cgi-bin/page/archive.pl?No=83
- (2) TerasIC TRDB-D5M Camera:
    https://www.terasic.com.tw/cgi-bin/page/archive.pl?Language=English&CategoryNo=68&No=281
- (3) USB to 6-Wire UART Cable
- (4) VGA Cable and Monitor (640 by 480)

For the complete codebase, please refer to the following link:
https://github.com/ShichenQiao/ECE554_SP23_FPGA_Handwriting_Recognition

For a video demonstration of our product, please refer to the following link:
https://youtu.be/7T7qIo2IxYQ

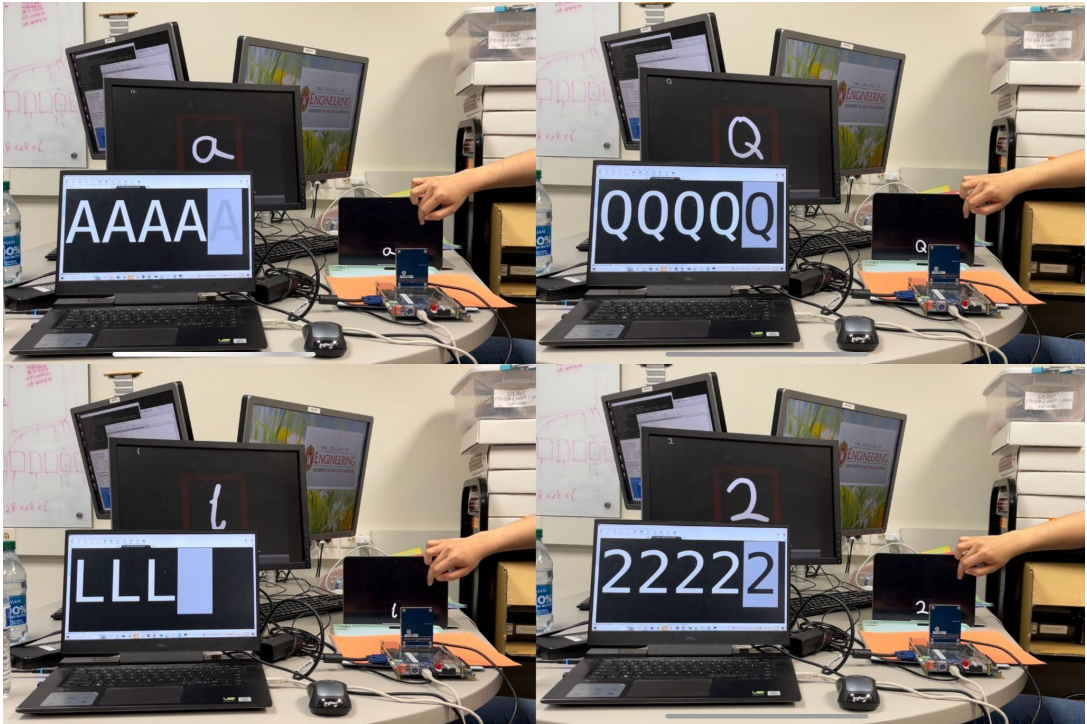

**Figure 1:** Selected screenshots from our video demo. They show the results of our FPGA-CNN model's prediction of several handwritten characters. The bottom-left screen shows the prediction result obtained, while the bottom-right corner features our FPGA board equipped with a camera. On the image's right-hand side, a hand can be seen swiping between sample handwriting on an iPad. Meanwhile, the monitor in the background displays the video feed captured by the camera. The red frame indicates the range of letter prediction. In the top-left corner of the monitor, a compressed 32x32 image of the target letter is echoed back.

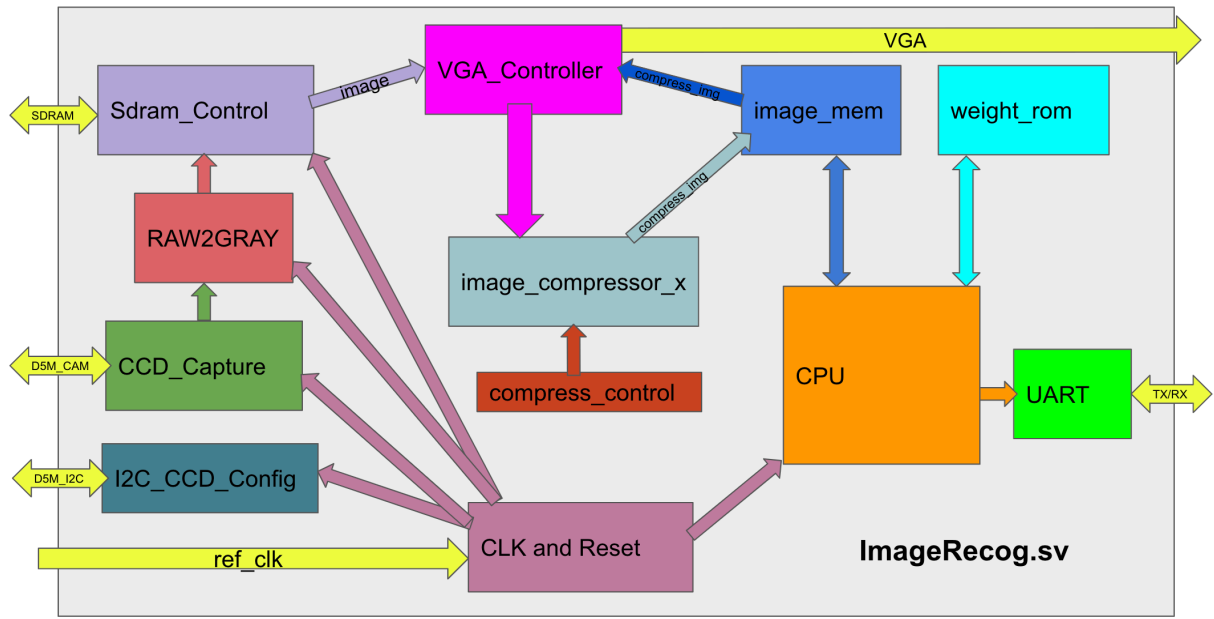

**Figure 2:** Top-level block diagram. The details of each module are elaborated in the following sections of this document.

## 2. Engineering Protocols

This section introduces the essential engineering standards we followed to help guide our ISA design, hardware implementation, UI development, and board-to-board communication. Understanding these concepts is necessary to replicate, configure and enhance our product.

### 2.1. IEEE 1800-2009 System Verilog

System Verilog is a hardware description and verification language used to design digital systems to improve productivity in verifying hardware designs [6]. It is an extension of Verilog, which includes additional features like assertions, constrained random testing, and coverage measurement. For example, we used System Verilog's casex feature to implement a part of the floating-point operations and the $bitstoshortreal system function to verify those designs. Some of our unit tests were also written in System Verilog.

### 2.2. IEEE-754 32-Bit Floating-Point Standard

IEEE-754 defines formats for representing and manipulating floating-point numbers, which are used to approximate real numbers in computers [1]. This project employs the single-precision 32-bit format to perform addition, subtraction, multiplication, and integer conversion on FP numbers. Hardware support for IEEE-754 32-bit FP arithmetic provides a broader range of representable values and greater accuracy than integer calculations.

### 2.3. Video Graphics Array (VGA)

The VGA is a display protocol designed by IBM in 1987 [2]. We utilized this protocol to echo live camera inputs and the screenshot under classification on a 640x480 monitor. The VGA_Controller module in Figure 2 above was used to control data transfers and timing. We feed our pixels directly from the SDRAM on the camera daughter card into the VGA Controller and reset the timing when the board is initialized. The VGA protocol operates at 25MHz.

### 2.4. Universal Asynchronous Receiver-Transmitter (UART)

Our project used the UART protocol to transmit predicted results for viewing [3]. To enable buffering of the transmitted data, we added a FIFO to a base UART and used it to send predicted characters to the display.

### 2.5. Inter-Integrated Circuit (I2C)

The I2C is a widely-used synchronous, multi-master/multi-slave (controller/target), packet-switched, single-ended, serial communication bus [4]. Our project used the I2C protocol to adjust the camera's exposure, pause, zoom, and brightness settings. Specifically, we used I2C_CCD_Config to change the camera settings. Once the camera captures an image, it is processed (using RAW2Gray) and stored directly in the SDRAM.

## 3. Instruction Set Architecture

The following ISA table documents all instructions our custom, five-stage, IEEE-754-compliant RISC processor supports:

Notes
1. a=opcode, c=sub_opcode, x=don't_care, d=destination, s=source, t=second_source, i=immediate, o=offset
2. Flag registers are Z-zero, V-overflow, N-negative/sign
3. The overflow flag denotes positive overflow as well as negative underflow
4. Register R0 is hard-wired to 32'h00000000, can't be written to
5. Jal instruction always stores the return address in register R31. Do not write R31 inside function calls if you wish to return.

| Instruction | Encoding | Sample Instruction | OPCODE | Sample Explanation | Other Comments |
|---|---|---|---|---|---|
| ADD | aaaa_axxx_xxxd_dddd_ xxxs_ssss_xxxt_tttt | ADD R1, R2, R3 | 5'b00000 | R1 <= R2 + R3 | Saturating arithmetic Updates the Z, V and N flag registers |
| ADDZ | | ADDZ R1, R2, R3 | 5'b00001 | R1 <= R2 + R3 only if Z=1 | |
| SUB | | SUB R1, R2, R3 | 5'b00010 | R1 <= R2 - R3 | |
| AND | | AND R1, R2, R3 | 5'b00011 | R1 <= R2 & R3 | Updates the Z flag register |
| NOR | | NOR R1, R2, R3 | 5'b00100 | R1 <= ~(R2 \| R3) | |
| SLL | aaaa_axxx_xxxd_dddd_ xxxs_ssss_xxxi_iiii | SLL R1, R2, C | 5'b00101 | R1 <= R2 << C | C is 5-bit unsigned immediate value Updates the Z flag register |
| SRL | | SRL R1, R2, C | 5'b00110 | R1 <= R2 >> C | |
| SRA | | SRA R1, R2, C | 5'b00111 | R1 <= R2 >>> C | |
| LW | aaaa_axxx_xxxd_dddd_ xxxs_ssss_oooo_oooo | LW R1, R2, O | 5'b01000 | R1 <= DataMem[R2 + O] | O is 8-bit signed immediate value |
| SW | | SW R1, R2, O | 5'b01001 | DataMem[R2 + O] <= R1 | |
| LHB | aaaa_axxx_xxxd_dddd_ iiii_iiii_iiii_iiii | LHB R1, C | 5'b01010 | R1 <= {C, R1[15:0]} | C is 16-bit signed immediate value |
| LLB | | LLB R1, C | 5'b01011 | R1 <= sign-extend{C} | |
| B | | | | | |
|    NEQ | aaaa_accc_xxxx_xxxx_ xxxx_oooo_oooo_oooo | B NEQ, label | 5'b01100000 | Branch if Z=0 | O is signed 12-bit offset in two's complement Branch target address = (Address of branch instruction + 1) + offset PC holds word addresses, each instruction is 1 word, offset is specified as the number of instructions with respect to the instruction following the branch instruction. |
|    EQ | | B EQ, label | 5'b01100001 | Branch if Z=1 | |
|    GT | | B GT, label | 5'b01100010 | Branch if {Z,N}==2'b00 | |
|    LT | | B LT, label | 5'b01100011 | Branch if N=1 | |
|    GTE | | B GTE, label | 5'b01100100 | Branch if N=0 | |
|    LTE | | B LTE, label | 5'b01100101 | Branch if N=1 or Z=1 | |
|    OVFL | | B OVFL, label | 5'b01100110 | Branch if V=1 | |
|    UNCOND | | B UNCOND, label | 5'b01100111 | Branch unconditionally | |
| JAL | aaaa_axxx_xxxx_xxxx_ xxxx_oooo_oooo_oooo | JAL label | 5'b01101 | R31 <= address of jal instruction +1, jump to target | O is signed 12-bit offset in two's complement Jump target address = (Address of jal instruction + 1) + offset |
| JR | aaaa_axxx_xxxx_xxxx_ xxxt_tttt_xxxx_xxxx | JR R31 | 5'b01110 | Jump to the address in R31 | Can be used to return from function calls (jal) |
| PUSH | aaaa_axxx_xxxx_xxxx_ xxxs_ssss_xxxx_xxxx | PUSH R1 | 5'b10010 | DataMem[SP] <= R1; Decrement SP | Stores value in R1 into data memory pointed by the stack pointer; decrements stack pointer |
| POP | aaaa_axxx_xxxd_dddd_ xxxx_xxxx_xxxx_xxxx | POP R1 | 5'b10011 | R1 <= DataMem[SP]; Increment SP | Loads value in data memory pointed by the stack pointer into R1; increments stack pointer |
| ADDI | aaaa_axxx_xxxd_dddd_ xxxs_ssss_iiii_iiii | ADDI R1, R2, I | 5'b10100 | R1 <= R2 + C | I is 8-bit signed immediate value. Updates the Z, V and N flag registers |
| SUBI | | SUBI R1, R2, I | 5'b10101 | R1 <= R2 - C | |
| MUL | aaaa_axxx_xxxd_dddd_ xxxs_ssss_xxxt_tttt | MUL R1, R2, R3 | 5'b11000 | R1 <= (signed) R2[15:0] * (signed) R3[15:0] | Only support 16 by 16 multiplications |
| UMUL | | UMUL R1, R2, R3 | 5'b11001 | R1 <= (unsigned) R2[15:0] * (unsigned) R3[15:0] | |
| ADDF | | ADDF R1, R2, R3 | 5'b11010 | R1 <= R2 + R3 (floating-point) | Floating-point operations 1-bit sign, 8-bit exponent, 23-bit mantissa ADDF, SUBF, MULF will set the flags for branch. |
| SUBF | | SUBF R1, R2, R3 | 5'b11011 | R1 <= R2 - R3 (floating-point) | |
| MULF | | MULF R1, R2, R3 | 5'b11100 | R1 <= R2* R3(floating-point) | |
| ITF | aaaa_axxx_xxxd_dddd_ xxxs_ssss_xxxx_xxxx | ITF R1, R2 | 5'b11101 | R1 <= R2 (integer to floating-point) | ITF and FTI will NOT set the flags for branch. |
| FTI | | FTI R1, R2 | 5'b11110 | R1 <= R2 (floating-point to integer) | |
| HLT | 1111_1xxx_xxxx_xxxx_ xxxx_xxxx_xxxx_xxxx | HLT | 5'b11111 | Processor Halt | For simulation and testing ONLY. |

For a Perl assembler implementation of this ISA, please refer to the following link:
https://github.com/ShichenQiao/ECE554_SP23_FPGA_Handwriting_Recognition/blob/main/Project/SourceCode/test/asmbl_32.pl

# 4. System Verilog Hardware Implementation

This section documents the usages, design ideas, and interfaces between our product's critical System Verilog hardware modules.

For implementation details, please refer to the .sv files in this directory:
https://github.com/ShichenQiao/ECE554_SP23_FPGA_Handwriting_Recognition/tree/main/Project/SourceCode

## 4.1. Top-Level

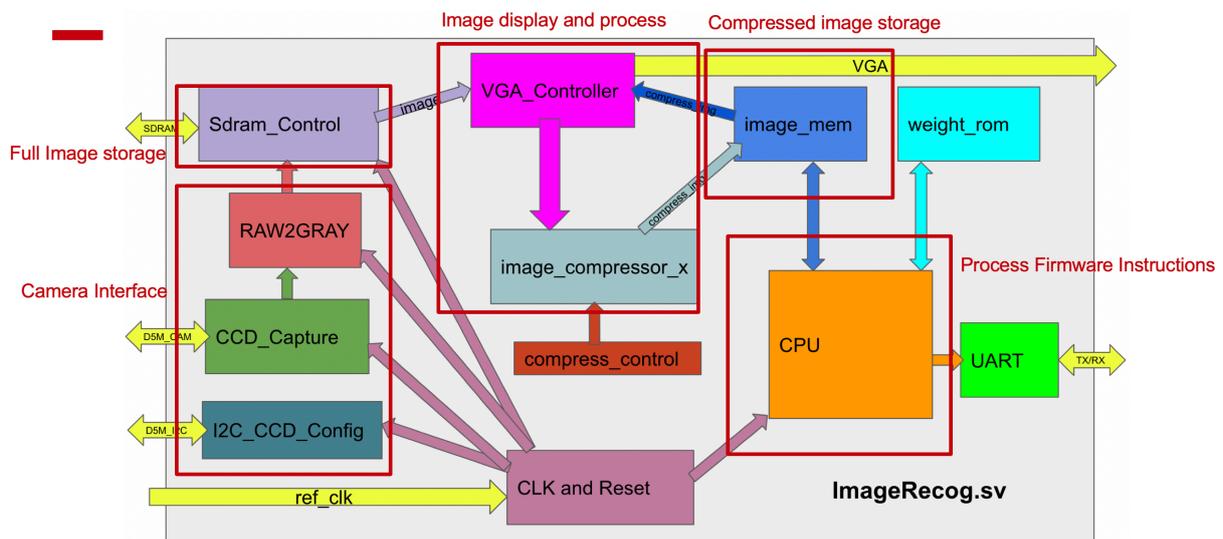

**Figure 2 (annotated):** Top-level block diagram. The details of each module are elaborated in the following sections of this document.

Link to code discussed in this section:
https://github.com/ShichenQiao/ECE554_SP23_FPGA_Handwriting_Recognition/blob/main/Project/SourceCode/ImageRecog.sv

At the top level, we utilized the following memory-mapped registers to control our peripheral I/O (LEDs, switches, and buttons) on the FPGA evaluation board, UART communication between the FPGA and the PC, and triggers for image captions:

| Register Address: | Description: |
|---|---|
| 0x0000C000 | Write to this address will write to LEDR[9:0] of board. |
| 0x0000C001 | Read from this address will return the state of SW[9:0] of board. |
| 0x0000C004 | UART Transmit Buffer (IOR/W=0); Receive Buffer (IOR/W=1). |
| 0x0000C008 | Set to 1 to request compressing the image, it will be knocked down once the compression is finished. |

In addition, two external memory modules are instantiated at the top level to support the implementation of CNN in hardware. The image_mem is a dual-port memory module to store a 32*32 compressed image. The weight_rom is a single-port memory module with all 63,654 weights necessary for CNN. These weights are trained and validated using the software. They are then loaded into the ROM using a weight.hex file generated by the software. We memory-mapped the weight_rom into the processor's address space.

The size and addresses of these SRAM blocks are documented in the following table:

| Name: | Start Addr | End Addr | Description: |
| --- | --- | --- | --- |
| image_mem | 0x00010000 | 0x0001030F | Image memory RAM, take inputs from image compressor and output values to VGA and processor. |
| weight_rom | 0x00020000 | 0x00021E9F | Weight memory ROM, values load from ML software, output to processor. |

Further, to implement real-time processing, we must be able to capture an image only after the last image is processed and outputted through UART. Therefore, we use compress control logic to allow the CPU to request a new image compression. Before processing each image and having the compressed snapshot in image_mem, the assembler code must request a snapshot by storing 1 to 0xC008 in data mem (SW 1, 0xC008). The compress control will wait until the SDRAM access is synchronized with the first pixel of the snapshot and enable the compressor to process the image. The assembler code needs to check the data in 0xC008 periodically. The compressed image is ready when the polled data in 0xC008 becomes 0.

Some of the critical control signals in our top-level module are noted in the following table:

| Signal: | Dir: | Description: |
| --- | --- | --- |
| uncompress_addr_x[7:0] | In | X axis of the uncompressed image address. It is used to synchronize with the first pixel of the uncompressed image. |
| uncompress_addr_y[7:0] | in | Y axis of the uncompressed image address. It is used to synchronize with the first pixel of the uncompressed image. |
| we | In | Write-enable signal for requesting a snapshot. The CPU will access this when writing to 0xC008. |
| compress_wdata | In | The request for a snapshot. The CPU can set this to 1 to indicate a request for a snapshot, 0 to indicate no need to take a snapshot. |
| pause | In | Input from the button. This supports the function of freezing video input by pressing a button (KEY[2]). The compress signal control will not start until the key is released. |
| compress_req | out | The status register of the compressor control. If it is 0, there is no compression going on. If it is 1, compression is in process. CPU is responsible for accessing 0xC008 to check this status. |
| compress_start | out | The control signal to start a new compression. |

## 4.2. Custom RISC CPU

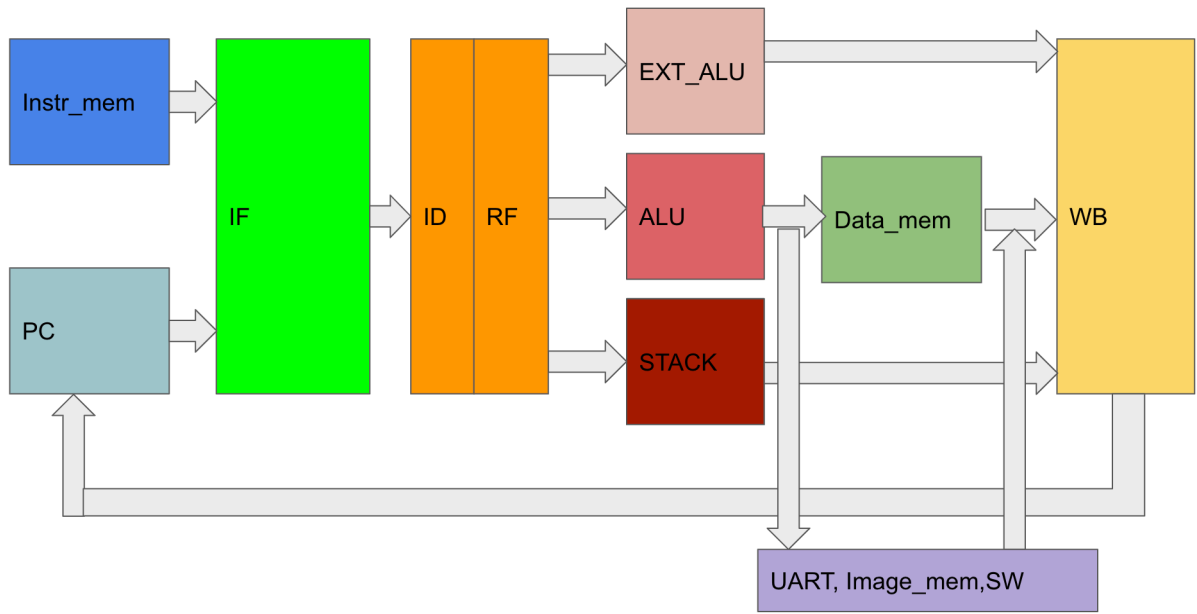

**Figure 3:** CPU block diagram.

Link to code (highest hierarchy) discussed in this section:
https://github.com/ShichenQiao/ECE554_SP23_FPGA_Handwriting_Recognition/blob/main/Project/SourceCode/cpu.v

Figure 3 above shows only a high-level block diagram of our 32-bit, 5-stage, floating-point RISC processor. Due to the limited block SRAM available on our FPGA, the Instr_mem only has 1K entries. PC stands for program counter and points to the index of the next instruction in Instr_mem. IF stands for the instruction fetch stage, and ID stands for the instruction decode stage. In the ID stage, a 32 by 32-bit register file (made from two dualPort32x32 SRAM units and bypass logic) is accessed, and branch controls are also managed. Further, the execute stage contains three major modules, ALU, EXT_ALU, and STACK. EXT_ALU contains compute units for all floating-point operations (which will be elaborated in the next paragraph) and integer multiplications; STACK contains a 1K SRAM First-In-Last-Out (FILO) buffer handling PUSH and POP instructions; ALU manages all other instructions in the scope of our ISA. In the memory stage, the CPU interacts with Data_mem (an internal SRAM block with 8K entries) and external memories (Image_mem as mentioned above, on-board switch registers, and UART). Finally, in the write-back (WB) stage, compute results are written back to RF when needed.

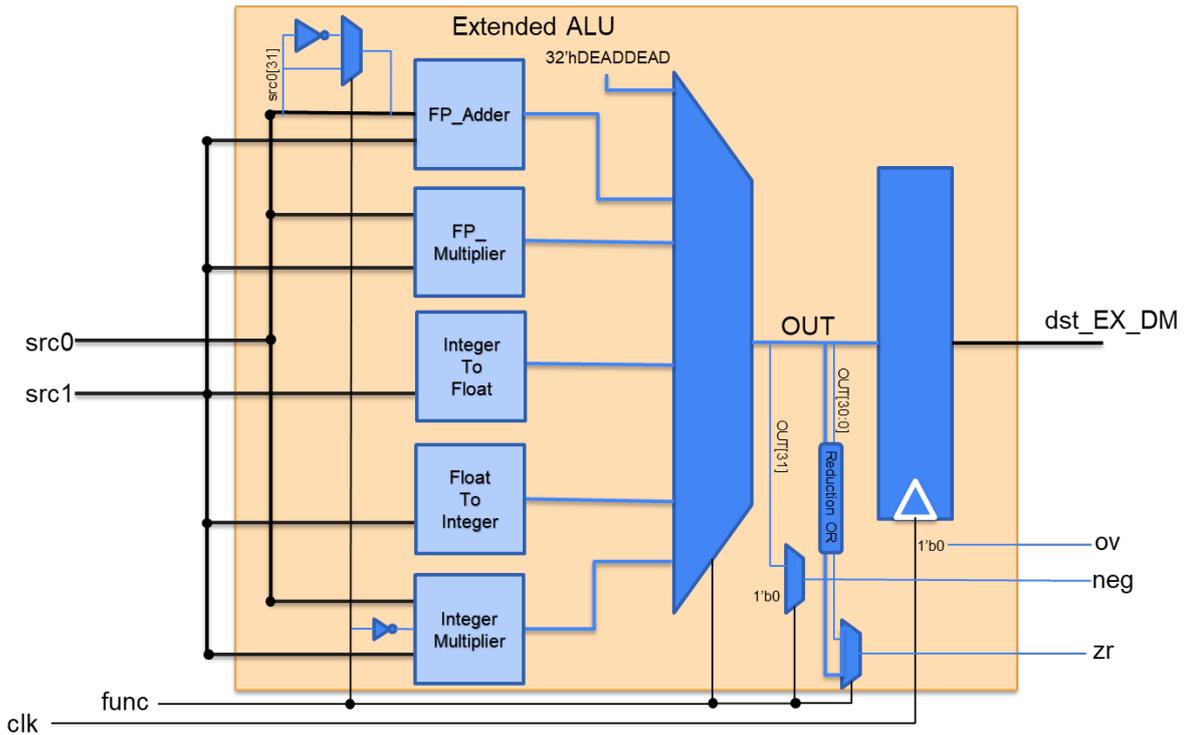

**Figure 4:** Extended ALU block diagram.

Figure 4 above shows EXT_ALU contains five submodules for floating-point addition and multiplication, conversions between float and integer, and integer multiplication. The func signal selects its output value and flags. A high-level interface is defined as follows:

| Signal: | Dir: | Description: |
|---|---|---|
| clk | in | 50MHz system clock |
| src1[31:0] | in | 32-bit source 1 into ALU |
| src0[31:0] | in | 32-bit source 0 into ALU |
| func[2:0] | in | 3-bit OP Code:<br>   000 ==> MUL<br>   001 ==> UMUL<br>   010 ==> ADDF<br>   011 ==> SUBF<br>   100 ==> MULF<br>   101 ==> ITF<br>   110 ==> FTI<br>   111 ==> undefined |
| dst_EX_DM[31:0] | in | 32-bit ALU output |
| ov | out | Overflow flag - but this is always 0!!! Kept here for following branch ops |
| zr | out | Zero flag - high when output is 0 (int zero or FP zeroes) |
| neg | out | Negative flag - high when output is negative |

Adding two IEEE-754 floating-point numbers is a complex process due to the potential difference in their exponents and unsigned mantissa. Here are the steps involved in the FP Adder process:

(1) Compare the exponents and determine the smaller ones. Calculate the absolute difference between the exponents; the larger exponent will be the common exponent.
(2) Prepend the common exponent to both mantissas, making them both 24-bit in length.
(3) Shift the mantissa with the smaller exponent to the right, using the lower 5 bits of the exponent difference as the shift amount. The maximum shift amount should be 22-bit.
(4) Convert both appended and shifted mantissa to 2's complement format. This makes both numbers 25-bit in length.
(5) Add the two 25-bit numbers, and if the result overflows positively or negatively. Note that this overflow is an internal overflow, not an external value overflow.
(6) Convert the 25-bit 2's complement result to a 25-bit signed number, where the MSB is the final sign, and the rest 24-bit is the unsigned value. If this 24-bit prepended mantissa starts with 0, but no internal occurs, denormalize the common exponent to 0.
(7) If an overflow occurs, shift the lower 24-bit to the right and append 1 in the MSB, or shift the lower 24-bit to the left if it has leading zeros. The common exponent, if not 0, is adjusted accordingly.
(8) The resulting mantissa is the lower 23-bit of the final 24-bit result, and the exponent is the final common exponent. The sign is the MSB of the final result.

Multiplying two 32-bit values in IEEE-754 format involves breaking the inputs into {S1, E1, M1} and {S2, E2, M2}. The signs are XORed to determine the sign of the product, and the exponents are added together with the 127 offsets accounted for. The mantissas are then appended with an implicit 1 (or 0) and multiplied. Special values like -INF, -0, +0, and +INF require special combinational logic to adjust the outputs. The resulting values are then concatenated and normalized to conform to the IEEE FP standard and output as a 32-bit value.

### 4.3. Camera Interface

This section discusses our camera interface modules (shown as SDRAM_Control, I2C_CCD_Config, Raw2Gray, and CCD_Capture in Figure 2). All code could be found with the same filenames in our source code directory.

To begin with, we need to understand the D5M camera interface ports [7]:

input: D5M_D[11:0], D5M_FVAL, D5M_LVAL, D5M_PIXCLK, D5M_STROBE
output: D5M_RESET_N, D5M_SCLK, D5M_TRIGGER, D5M_XCLKIN
inout:  D5M_SDATA

These ports are implemented using GPIO_0 ports on the FPGA board. We map the ports on GPIO to the corresponding pins on the D5M camera. The D5M camera captures the images, and the data is stored in the external SDRAM on the FPGA developer board. Because we do not have sufficient FPGA-embedded SRAM memory, we must use the external SDRAM.

To manage the data transferred into the external SDRAM, we use the Sdram_Control module to manipulate the timing and data sent into the SDRAM. Each received pixel is divided into two parts because the width of the SDRAM memory is 8 bits, and the received pixel is 12 bits. The Sdram_Control module takes the captured image and sequentially stores the upper and lower bytes. The Sdram_Control module has a FIFO buffer to avoid loss of data.

To manage the Camera's configuration, we use the I2C_CCD_Config module to adjust the exposure, zoom, and brightness configuration of the D5M camera through the I2C protocol. When we want to adjust one camera setting, we use the combination of a switch and a button to set the desired setting. For example, if SW[0] is on and KEY[1] is pressed, the exposure will increase. If SW[0] is off and KEY[1] is pressed, the exposure will decrease. I2C_CDD_Config will monitor the switch and keys to update settings through the I2C protocol.

To convert the captured color image into a grayscale image, we use the Raw2Gray module to process the captured pixel and convert it to a grayscale image for CNN prediction. The Raw2Gray module takes the input from the CCD_Capture module, which is captured from the camera and converts the output to grayscale values.

To capture the image and manage the communication protocol with the camera, we use the CCD_Capture module to control the clock and the data received. This module also keeps track of the received pixel's frame count, x_location, and y_location.

As a result, whenever the firmware provides a valid image capture request signal through the memory-mapped registers, an image capture cycle will be triggered, and a raw image will be stored in the external SDRAM.

### 4.4. Image Processing Units
**Note:** In this section, this blue color indicates CNN-only implementations (which will be elaborated on in section 5.3).

Link to code discussed in this section:
https://github.com/ShichenQiao/ECE554_SP23_FPGA_Handwriting_Recognition/blob/main/Project/SourceCode/image_compressor_x.sv

The raw image taken from the camera is first stored in SDRAM and then sequentially fed into the image compressor. This image compressor sequentially takes a 224*224 8-bit image from SDRAM and compresses it into a 28*28 8-bit image by taking the average color among 8*8 blocks, which is sufficient for our LC and NN models (see sections 5.1 and 5.2, respectively). Note that the image compressor processes only one pixel in each CPU cycle.

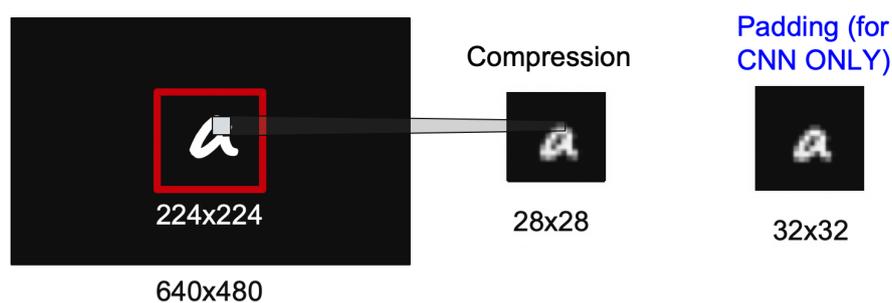

**Figure 5:** Sample image processing for our CNN model.

The image compressor x is designed explicitly for the CNN models, requiring a 32*32 image with zero-padding of width 2 surrounding the original 28*28 image. The image compressor x, therefore, has a different output signal called compress_addrx, which ranges from 0 to 1023 (instead of 0 to 783). The zero-padding is achieved by skipping and not writing to the padding address in the image memory so that these addresses always contain 0. The compress_addrx signal is a combinational logic of the original output compress_addr.

The high-level interface and register usages of these image-processing units are defined below:

| Signal: | Dir: | Description: |
|---|---|---|
| clk | in | 25MHz clock signal from VGA display |
| rst_n | in | System reset signal |
| start | in | Signals a valid pixel color input starting from 0 |
| pix_color_in[7:0] | in | 8-bit pixel color value from VGA DRAM (0 to 255 grayscale) |
| pix_haddr[7:0] | in | 8-bit pixel horizontal address (0 to 223) |
| pix_vaddr[7:0] | in | 8-bit pixel vertical address (0 to 223) |
| sram_wr | out | Write enable signal to the SRAM memory storing compressed image |
| pix_color_out[7:0] | out | 8-bit compressed pixel color by taking average value among an 8*8 block |
| compress_addr[9:0] OR compress_addrx[9:0] | out | 10-bit compressed image pixel address (0 to 783 for a 28*28 image) OR 10-bit compressed image pixel address (0 to 1023 for a 32*32 image) |

| Register: | Description: |
|---|---|
| compress_addr[9:0] | Described in table above.<br>Reset to 10'd784; zero-set when start asserted; incremented when sram_wr asserted.<br>Namely, it increments to the next available SRAM address after an image memory write and stalls itself when the entire image memory gets written until the next asserted start signal. |
| block[13:0][0:27] | 28 14-bit wide SRAM blocks to store the accumulated sum of every pixel value inside 28 8*8 blocks.<br>We need 28 of them since at least one row of 8*8 blocks should be saved for averaging, and there are 28 blocks per row (224/8 = 28). Its address is determined by the upper 5 bits of pix_haddr, named b_haddr.<br>14-bit is needed since it stores the sum of 64 8-bit wide pixel color values. The average value is taken from its upper 8 bits to produce a compressed pixel color value. |

# 5. Training Software

As mentioned in the summary section, three different ML architectures were implemented and validated on our product: linear classification, a 784-64-10 fully connected neural network (NN), and a LeNet-like Convolutional Neural Network (CNN) with ReLU activation layers and 36 classes [5]. To reiterate, we fully manage the classification processes on the FPGA, but the training steps are completed on PC in advance. The details of how we trained the models and obtained the kernels and weights are shared in this section. Note that the final product only utilizes the CNN model due to LC and NN performance issues.

Link to the Python Notebook for CNN training:
https://github.com/ShichenQiao/ECE554_SP23_FPGA_Handwriting_Recognition/blob/main/Project/ML/train.ipynb

## 5.1. Linear Classification Model

| Input: | Output: |
|---|---|
| keras.datasets.emnist 120,000 images with labels | weight.hex contains 28,224 8-digits hex numbers, one per line. |

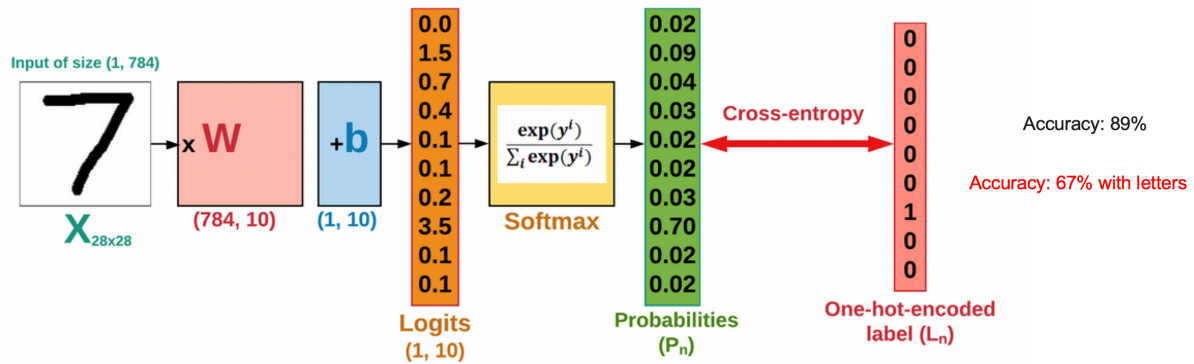

**Figure 6:** Linear Classification Model [9].

This software was written in Python on Google Colaboratory, and it works in three steps: reading train data, training on the data, and producing the weight matrix. The Keras EMNIST dataset contains 60,000 different images of 28*28 pixels of handwritten numbers and corresponding labels to indicate their values [8]. The first 40,000 images were used for training, the following 10,000 images were used for validation during training, and the last 10,000 images were used for testing the accuracy of the trained model.

More specifically, this model is a softmax loss linear classifier, which utilizes the difference between ideal logistic probabilities and the current logistic probabilities as the loss function and employs gradient descent to minimize the loss function. The model was trained for 200 epochs using the PyTorch library. With the appropriate learning rate and regularization factors, the training accuracy was over 90%, and the testing accuracy was around 88.6% for recognizing only digits. However, with the letters also concerned, the accuracy was only 67%.

After finding the best-performance, digits-only model, we extracted the transpose of the weight matrix. The transposed weight matrix was flattened into a list of 784*36 = 28,224 floating-point numbers (7,840 for digits-only models). Each row of the 28 numbers was parsed into hex numbers and pasted into a .hex file, where each line was in the format of {LINE_NUMBER} {HEX_NUMBER}. This file can be directly loaded into the FPGA board as a ROM and later read for matrix multiplications handled by the firmware.

## 5.2. Neural Network Model

| Input: | Output: |
|---|---|
| keras.datasets.emnist 425,600 images with labels | weight_nn.hex contains 52,480 8-digit hex numbers, one per line |

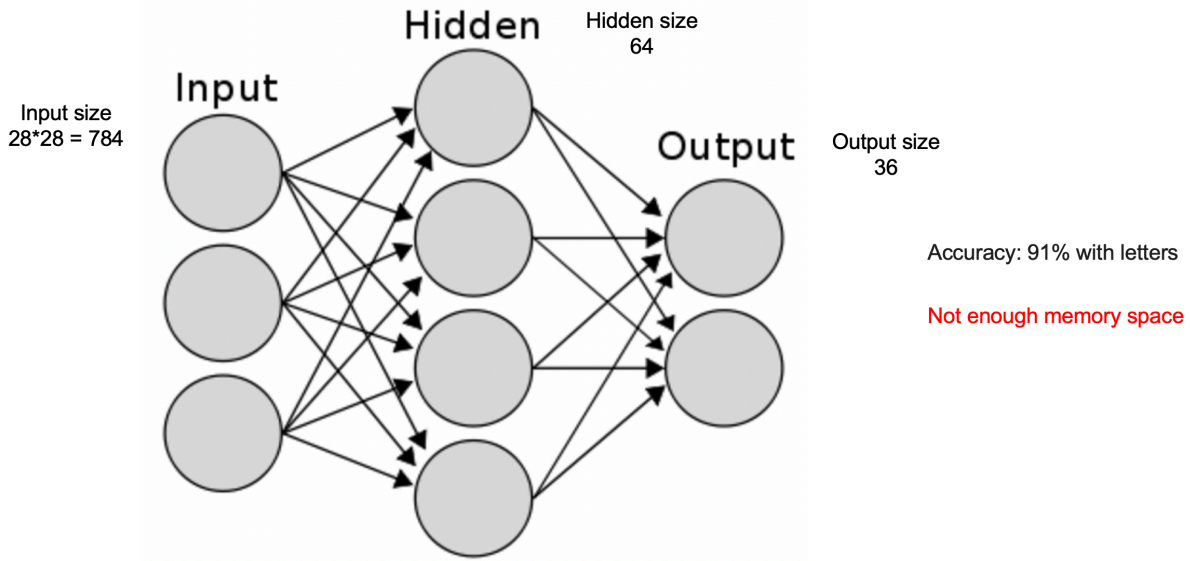

**Figure 7:** Neural Network Model [10].

Since Linear Classification didn't perform well, with around 89% accuracy for digits only and 67% accuracy for digits and letters, we changed our model to Neural Network. A Neural Network (NN) model typically has better performance in classification tasks when compared to Linear Classification (LC) models. This model is also written with Python but using Jupyter Notebook instead since we localized the training dataset.

We used the PyTorch library for this NN model with two neuron layers: 784*64 and 64*36, and there was a ReLU hidden layer between these two layers. We set these configurations for the torch.nn library and let it train/test for us. For predicting each image, this model first performs matrix multiplication with the image data and the first layer, 1*784 dot 784*64, resulting in a matrix of 1*64. Then, it changes all the negative numbers in this 1*64 matrix to 0. This simulates the ReLU layer in our NN model. Next, we will perform another matrix multiplication between 1*64 and 64*36. This results in 1*36 numbers, representing the score for each class (10 for digits and 26 for letters - both upper-case letters and lower-case letters classify into the same category). The final result is around 99% accuracy for digits only and 91% for digits and letters. However, when both letters and digits are concerned with this model, the weight hex file is so large that it cannot be stored in the SRAM of our DE1 FPGA.

We exported the two neuron layers to extract the weights for the digits-only model. model.lin1 and model.lin2 were printed to the weightnn.hex file, one on each line. model.lin1 has dimension of 784*64 = 50,176, and model.lin2 has dimension of 64*36 = 2,304. This totals up to 50,176+2,304 = 52,480. This hex file will later be read into weight_nn_rom.sv. Each number is first converted to a float, then cast into a hex, and then re-structured into an 8-digit hex num that is 32 bits large.

## 5.3. Convolutional Neural Network Model

| Input: | Output: |
|---|---|
| keras.datasets.emnist  425,600 images with labels | weight_cnn.hex  contains 63,654 8-digit hex numbers, one per line. |

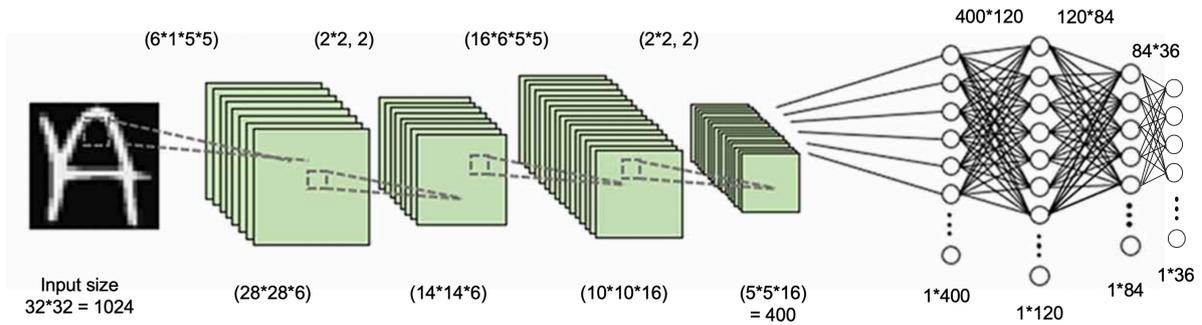

**Figure 8:** Convolutional Neural Network Model [11].

The NN model was much better than the CL model in terms of accuracy, but we weren't satisfied with its performance on our FPGA. Having only two layers limits the model's real-time accuracy performance. But if we were to have more layers for the NN model, our FPGA would have run out of memory, as we need to store the weight information on the board. Another model introduced is the Convolutional Neural Network (CNN), which would shrink the input size by only extracting critical features from the dataset and then sending the convoluted data to NN layers. This helped us to keep the total number of weights manageable.

For the CNN model, we kept most of the structure of the NN model but added more convolutional layers before the linear layers. Two convolutional layers are added, each of their kernels having the dimensions of 6*1*5*5 and 16*6*5*5, which works like two filters for the input data. The input data is still 784 (28*28*1). The first convolutional layer will add paddings of 2 pixels on each of the four sides of this picture, which will be 32*32*1. The padding added will all have a value of 0. This 32*32*1 matrix will be sent to the model.cov1 layer, which uses 6*1*5*5 kernels to extract information. This operation will then use a 28*28*6 matrix as the first, middle data. We then change all negative numbers to a 0, serving as the ReLU layer. This processed 28*28*6 data will then be average pooled with 2*2 kernels and stride = 2, which means each time we take a 2*2 matrix and grab the average, and move by two each time. This will then give us a 14*14*6 matrix as the input to the second convolution layer. The model.cov2 has 16*6*5*5 kernels, and by operating on the 14*14*6 matrix, which will turn into a 10*10*16 matrix as the second middle data. Again, we take all negative values out and substitute them with 0. Then we do average polling again with 2*2 kernels and a stride of 2. This will finally give us a 5*5*16 matrix as our data. The data has a dimension of 1*400 after being flattened, which is significantly less than the original input of 1*784. The total number of weights for the convolution layers is the sum of the two convolutional layers, in which model.conv1 is 6*1*5*5 = 150, and model.conv2 is 16*6*5*5 = 2400. The dimension of the two convolutional layers is 150 + 2,400 = 2,550 numbers.

This 1*400 matrix will then be sent to the NN model, which has three layers: 400*120, 120*84, and 84*36(84*10 for the digits-only model). The total numbers of weight for the NN portion is the sum of these three layers. model.lin1 has 400 * 120 = 48,000, model.lin2 has 120 * 84 = 10,080, and model.lin3 has 84*36 = 3,024. In total, the dimension of the three linear layers is 61,104. The total dimension used by weight_cnn_rom.sv will be 61,104 + 2,550 = 63654. For the digits-only model, the total number of weights would be 48,000 + 10,080 + 840 + 2,550 = 61,470. These numbers entail that 250KB of SRAM is sufficient to store all kernels and weights for this CNN model, which fits well on our FPGA.

Eventually, this model also demonstrated the highest accuracy in simultaneously classifying upper-case letters, lower-case letters, and digits, according to the statistics returned from our test set by PyTorch.

## 6. Embedded Firmware

From the firmware's perspective, the input image (integers between 0 and 255) is stored in image_mem starting at 0x00010000, and the pre-trained kernels or weights (in 32-bit FP format) are stored in weight_rom (or weight_nn_rom or weight_cnn_rom) starting at 0x00020000. With help from the same machine learning model as the training processes described above, the firmware would accomplish the classification processes through programs written in our custom assembly language.

We chose to only explain the main function in CNN_supercharged.asm in this paper, as it's the most complicated one compared to the main functions among all the firmware we developed.

Link to the assembly code discussed in this section:
https://github.com/ShichenQiao/ECE554_SP23_FPGA_Handwriting_Recognition/blob/main/Project/firmware/CNN_supercharged.asm

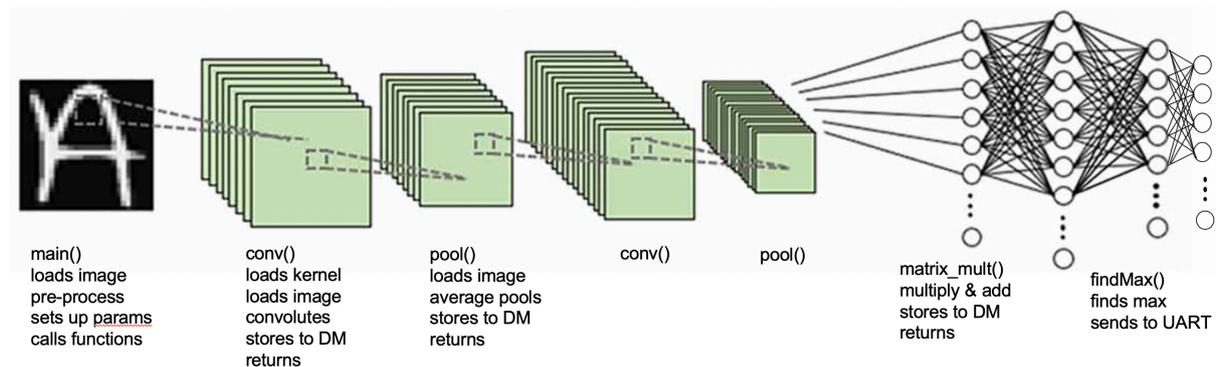

**Figure 9:** CNN_supercharged.asm code summary [11].

### 6.1. Main Function

The main function is an infinite loop, tailoring the different layers together through data memory (DM) and indefinitely triggering CNN classifications. Before entering a CNN, MAIN sends a snapshot request to the image processing unit and waits in SNAPSHOT_WAIT until an image is captured from the camera, compressed, padded, and stored in image_mem. Then MAIN would call PRE_PROCESS to convert the integers in image_mem to 32-bit floating-point format and store them in DM.

After pre-processing, MAIN setup parameters such as a pointer to weight ROM, input matrix size, channel lengths, DM pointer to output matrix, and so on for the different layers of the CNN and call CONV, AVG_POOL, MATRIX_MUL, and OUTPUT_LAYER in the order as shown in Figure 9 above to perform classifications. The DM and RF usage are listed below:

| Starting Addr | Ending Addr | Usage in this Program |
|---|---|---|
| 0 | 1023 | Preprocessed input image after HW padding (2 on each side) |
| 1024 | 5727 | Output of first convolution layer |

| 5728 | 6903 | Output of first pooling layer |
| (re-usage start) | | |
| 0 | 1599 | Output of second convolution layer |
| 1600 | 1999 | Output of second pooling layer |
| 2000 | 2119 | Output of first full NN layer |
| 2120 | 2203 | Output of second full NN layer |
| (re-usage end) | | |
| 7000 | 7399 | Work zone for matrix multiplications |
| 8000 | 8009 | Final Scores of the ten classes |

| Register Name | Usage in this Function |
|---|---|
| R0 | hard-wired 0 |
| R1 | snapshot trigger |
| R2 | pointer to weight ROM |
| R3 | pointer of image MEM |
| R4 | input matrix size |
| R5 | output matrix size |
| R6 | convolution input channel length |
| R7 | convolution output channel length |
| R27 | Snapshot status |
| R28 | reserved for result of matrix multiplication |
| R29 | DM pointer to output matrix |
| R30 | 0x0000C000 base address of peripherals |
| R31 | reserved for JAL/JR |

## 6.2. Pre-Processing

The PRE_PROCESS function is straightforward - it takes in a starting address and a matrix size as parameters, reads out each entry, converts each integer to FP format, and stores the results back to DM starting at address 0.

| Params: | |
|---|---|
| R3 | pointer of image matrix |
| R4 | matrix size |
| Local Variables: | |
| R0 | hard-wired 0 |
| R5 | DM pointer |
| R6 | temp reg holding value being converted |
| R31 | reserved for JAL/JR |

## 6.3. Convolution Layer

Our ML architecture uses a 5x5 convolution kernel in each convolution layer. To perform the convolution with a 5x5 kernel, we need to perform 5x5x#input_channels to get a pixel for the output. We need to generate number-of-output-image pixels depending on the output size and the number of output_channel. For example, if the input is 6 of 32x32 images (6 channels) and we want to output 16 images, then the output is 16 of 28x28 images, and we need to perform 16x28x28x5x5x6 multiplications. We also need to perform 16x28x28x5x5x6 additions to finish this convolution layer.

The CONV layer is configurable through its parameters - the input kernel's starting address, the image's starting address, the input image's size, the input channel length, the output channel length, and the output address. The register usage is defined below:

| Params: | |
|---|---|
| R2 | addr_kernel (start address of kernel) |
| R3 | addr_image (start address of image) |
| R4 | side_length_input (input side_length) |
| R6 | in_channel_length (repeat the convolution calculation for multiple images with the same kernel) |
| R7 | out_channel_length (repeat the convolution calculation for same image with different kernels) |
| R29 | addr_output (start address of output address, increase by one when a pixel is calculated) |
| **Local Variables:** | |
| R4 | side_length_output, same reg as input (will set to side_length_input - 4 to reflect the output side_length) |
| R5 | Set to R4 - 1 for branch purposes |
| R8 | x_result, x location of output images ( start as 0, increase by one once a pixel is calculated. set to 0 when reaching side_length_output) |
| R9 | y_result, y location of output images ( start as 0, increase by one when x_result reaches side_length_output ) |
| R10 | pix_sum (sum of result at a result_pix) |
| R11~R15 | 5 weight registers (shared between all 25 weights) |
| R16~R20 | 5 pix registers (shared between 25 pixels, it also stores the mult result) |
| R21 | image_length (use this jump distance to switch between input channels) |
| R22 | base (a temp base location for pixel load) |
| R23 | temp (intermediate for base address calculation |
| R24 | channel_id (a down counter for keep track of the channel id in process) |
| R25 | side_length_output, same reg as input (will set to side_length_input - 4 to reflect the output side_length) |
| R26 | static copy of param out_channel_length |

### 6.4. Average Pooling Layer

The average pooling layer comes after every convolution layer in our CNN. This layer has a function similar to the image compressor but is performed by firmware, which gives us more flexibility over other ML architectures – we only need to modify the assembly code if we want a different pooling kernel size or a different stride. The current code takes a 2*2 pixel block from an image specified by the layer starting address, calculates the average value, and stores it back to data memory as a compressed new image. It performs the steps above for images in every channel and generates new images of the number of channels. This layer aims to reduce the computations required while preserving the features of the image. Compressing the image also helps to prevent overfitting and improves the network's generalization ability.

| Params: | |
|---|---|
| R3 | layer starting address |
| R4 | image width |
| R6 | number of image channels |
| R29 | output starting address |

| Local Variables: | |
|---|---|
| R0 | hard-wired 0 |
| R2 | 0.25F |
| R7~R10 | pooled pixel |

## 6.5. Neural Network (Matrix Multiplication) Layer

Fully connected NN can be abstracted to loops of 1D matrix multiplications. Thus, we designed this MATRIX_MUL function which takes two starting addresses, one pointing to the weight and the other pointing to the image (or input data of a middle layer), multiplies the two one-dimensional (flattened) matrices together, and returns the accumulated value. The caller would record and manage the requested input and output dimensions and reconstruct the output matrix. Unlike the functions above, this function does not directly manipulate the DM. Instead, it returns the result to the caller function in R28. MAIN organizes these return values.

| Params: | |
|---|---|
| R2 | pointer to weight matrix |
| R3 | pointer of image matrix |
| R4 | matrix size |
| **Return Value:** | |
| R28 | result of matrix multiplication |
| **Local Variables:** | |
| R0 | hard-wired 0 |
| R5 | intermediate mult result store address |
| R6 | image pixel value |
| R7 | weight value |
| R8 | multiplication result |
| R31 | reserved for JAL/JR |

## 6.6. Output Layer

The output layer is responsible for making a prediction based on the input data. There are 36 possible outputs in our case, and the output layer determines the maximum result among all the possibilities. To achieve this, the layer loops through all 36 results and compares them with the current maximum value, replacing the current maximum when necessary. Once the maximum value is found, the output layer transmits the final prediction to the UART module. The prediction is in the form of an ASCII value that represents the predicted character.

| Params: | |
|---|---|
| R29 | base pointer to the output layer |
| **Local Variables:** | |
| R0 | hard-wired 0 |
| R2 | loop index i |
| R3 | loop terminate condition = 35 |
| R4 | current number |
| R5 | current number - current max |
| R6 | 35 - i |
| R7 | current max |
| R8 | current max index |
| R9 | 0x00000030 ASCII number offset |
| R10 | temp reg to distinguish between digits and letters |

| R30 | 0x0000C000 base address of peripherals |
| R31 | reserved for JAL/JR |

# 7. Validation

We started with System Verilog unit tests to validate our design, primarily focused on the floating-point computing and image processing units. We then validated each supported instruction in our processor by a Python script running VSIM commands. Finally, we created arbitrarily chose some fixed handwriting images, ran them through the PyTorch classifier with our trained model, dumped intermediate values from each ML architectural layer, and compared them to the simulated DM values in ModelSim. Passing these massive tests entails the correctness and accuracy of our system compared to our PyTorch software. The details are shared in this section.

## 7.1. Unit Tests in System Verilog

We validated its stand-alone correctness before integrating any System Verilog module into the Source Code folder. The IEEE-754 Floating-Point compliance tests are the most worth noting among all the unit tests we implemented in System Verilog.

We defined these corner case values according to the 32-bit Floating-Point Standard:

```
1.  localparam [31:0] FP_POS_SUB_MIN = 32'h0000_0001;
2.  localparam [31:0] FP_POS_SUB_MAX = 32'h007F_FFFF;
3.  localparam [31:0] FP_POS_MIN = 32'h0080_0000;
4.  localparam [31:0] FP_POS_MAX = 32'h7F7F_FFFF;
5.
6.  localparam [31:0] FP_NEG_SUB_MIN = 32'h807F_FFFF;
7.  localparam [31:0] FP_NEG_SUB_MAX = 32'h8000_0001;
8.  localparam [31:0] FP_NEG_MIN = 32'hFF7F_FFFF;
9.  localparam [31:0] FP_NEG_MAX = 32'h8080_0000;
10.
11. localparam [31:0] FP_SLT_ONE = 32'h3F80_0001;
12. localparam [31:0] FP_LST_ONE = 32'h3F7F_FFFF;
13. localparam [31:0] FP_SLT_NEG_ONE = 32'hBF7F_FFFF;
14. localparam [31:0] FP_LST_NEG_ONE = 32'hBF80_0001;
15.
16. localparam [31:0] FP_POS_ZERO = 32'h0000_0000;
17. localparam [31:0] FP_NEG_ZERO = 32'h8000_0000;
18. localparam [31:0] FP_POS_INF = 32'h7F80_0000;
19. localparam [31:0] FP_NEG_INF = 32'hFF80_0000;
20.
21. function bit is_NaN(input logic [31:0] FP_val);
22.     is_NaN = &FP_val[30:23] && |FP_val[22:0];
23. endfunction
```

Then, we tested all combinations of these values with other arbitrary non-corner-case values and fully random values with loops and the $bitstoshortreal and $shortrealtobits built-in functions in System Verilog.

For instance, these loops are used to validate ADDF and MULTF instructions:

```
1. for(int i = 0; i < 16; i++) begin
2.     for(int j = 0; j < 16; j++) begin
```

```
3.          A = SPECIAL_VALS_ARR[i];
4.          a = $bitstoshortreal(A);
5.          B = SPECIAL_VALS_ARR[j];
6.          b = $bitstoshortreal(B);
7.          o = shortreal'(a * b);
8.          product = $shortrealtobits(o);
9.  ...
```

```
1.  for(int i = 0; i < 1000000; i++) begin
2.          #1;
3.          a = $random();
4.          b = $random();
5.          as = $bitstoshortreal(a);
6.          bs = $bitstoshortreal(b);
7.          os = as + bs;
8.          sum = $shortrealtobits(os);
9.  ...
```

We only moved on to upper-level tests after all unit tests were passed. The complete set of unit tests could be retrieved from these directories:
https://github.com/ShichenQiao/ECE554_SP23_FPGA_Handwriting_Recognition/tree/main/Project/FP_modules
https://github.com/ShichenQiao/ECE554_SP23_FPGA_Handwriting_Recognition/tree/main/Project/other_extended_modules

## 7.2. Automatic ISA Validator in Python

Link to the tester code discussed in this section:
https://github.com/ShichenQiao/ECE554_SP23_FPGA_Handwriting_Recognition/blob/main/Project/SourceCode/test.py

Link to all self-checking assembly test cases discussed in this section:
https://github.com/ShichenQiao/ECE554_SP23_FPGA_Handwriting_Recognition/tree/main/Project/SourceCode/test

The Python auto-tester works together with the cpu_tb.sv file and the asm tests. The asm tests are designed to be stuck at 0x00AD if they pass and at 0x00DD if they fail. The cpu_tb.sv will wait for 300 cycles (which is more than most tests), then it will check if the PC is around 0x00AD or 0x00DD and output a message of "Test pass" or "Test fail."

The Python auto-tester has four functionalities: help, run one file, run all files, and clean:
"*python test.py help*" will display a help message on how to use the tester.
"*python test.py <filename>.asm*" will look for that .asm file in the test directory and assemble, compile, and simulate it.
"*python test.py*" will look for all .asm files in the test directory and assemble, compile, and simulate them.
"*python test.py clean*" will clean all .hex files in the test folder.

This Python script will search for all .asm files in the *test* directory, compiles them into .hex files, put the name of one program that has yet to be tested into instr_mem.sv, and then (re)compiles all .v and .sv files. Next, the script simulates the project and stores the output of the cpu_tb.sv into an output file. cup_tb.sv checks the pc and sees where it is stuck. If the PC is stuck around the passing loop, it will display a "test passed" message; otherwise, it will display "test failed." The test.py script will check if the output file test.output contains "pass," then we can remove the output file and proceed to the next asm file. If the file contains "fail," the script adds information about which test failed in the output file and blocks further testing. After all the testing procedures are finished, the test script will restore instr_mem.sv with its original content and display the testing results.

After passing all these tests, we were confident that implementing our ISA was good enough for full-chip/system validations.

## 7.3. Top Level Validation in ModelSim

Finally, we validated our top-level design with ModelSim simulations. The intermediate matrices and final scores reported by simulations were compared to their counterparts generated by our ML software using the same fixed image hex files. Note that with fixed hex images, our design takes about 3.7 million simulation cycles to classify one character. Since our software utilizes 64-bit floating-point operations, but our product uses a 32-bit format, minor errors at the least significant bits (when represented in decimal format) were allowed. Some example Python code we used to dump weight matrices, fixed images, and intermediate matrices are shared below:

*Weight Matrix Dump:*
```
1. write_file = open('test_weightfix.hex', 'w')
2. for i in range(len(model.conv1.weight.reshape(-1))):
3.     s_print = "@"+hex(i)[2:].zfill(4)+" "+hex(struct.unpack('<I',
   struct.pack('<f', model.conv1.weight.reshape(-1)[i].item())))[0])[2:].zfill(8)
4.     write_file.write(s_print.strip()+'\n')
5. write_file.close()
```

*Image Data Dump:*
```
1. write_file = open('test_fixed_image_cnn.hex', 'w')
2. for i in range(len(comb_test[sample_idx][0].reshape(-1))):
3.     s_print = "@"+hex(i)[2:].zfill(4)+"
   "+hex(int(comb_test[sample_idx][0].reshape(-1)[i]*255))[2:].zfill(2)
4.     write_file.write(s_print.strip()+'\n')
5. write_file.close()
```

*Prediction Data Dump:*
```
1. write_file = open('test_con1.txt', 'w')
2. for i in activation['conv1'].reshape(6*28*28):
3.     write_file.write(str(i.item())+"\n")
4. write_file.close()
5. write_file = open('test_pool1.txt', 'w')
6. for i in
   F.avg_pool2d(F.relu(activation['conv1']),kernel_size=2,stride=2).view(-1):
7.     write_file.write(str(i.item())+"\n")
8. write_file.close()
```

We compared selected sets of the intermediate layers and the final 36 scores of the classes. We validated that our hardware design precisely executed the classification processes for our different ML architectures. Selected screenshots of our design performance and validation results are shared below (Figures 10 through 13):

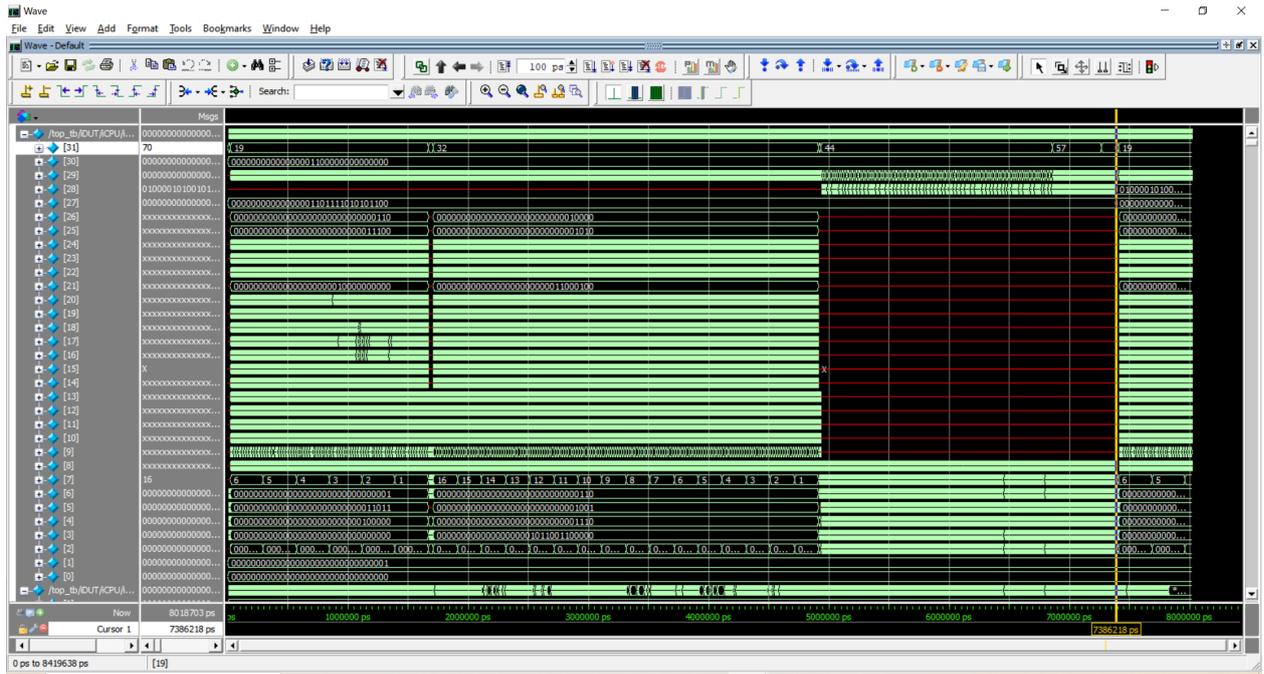

**Figure 10:** Register usage of a complete CNN classification process.

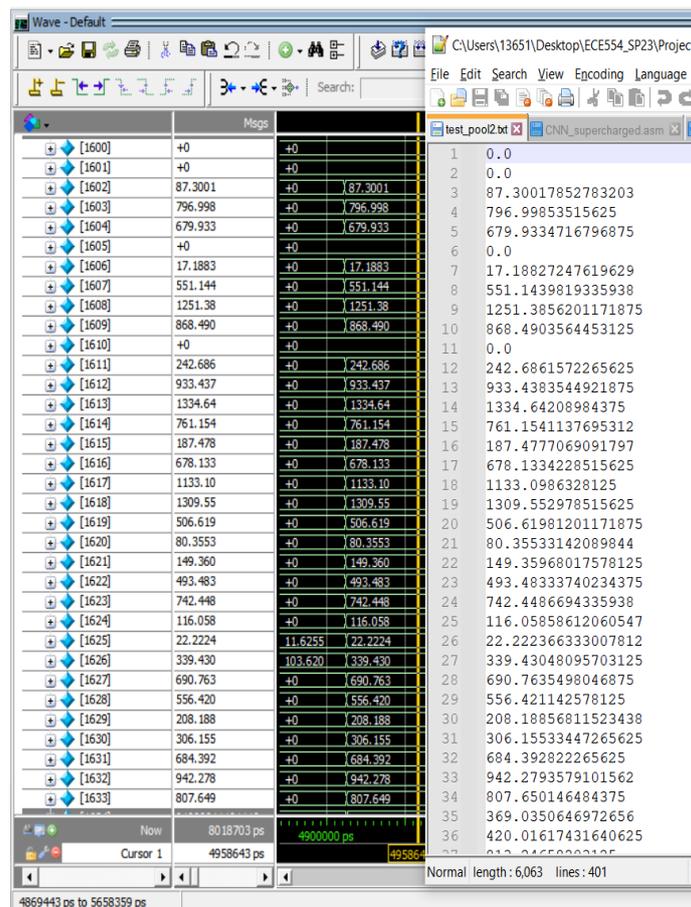

**Figure 11:** Intermediate layer validation - comparing FP values in the DM of our processor to software dumps.

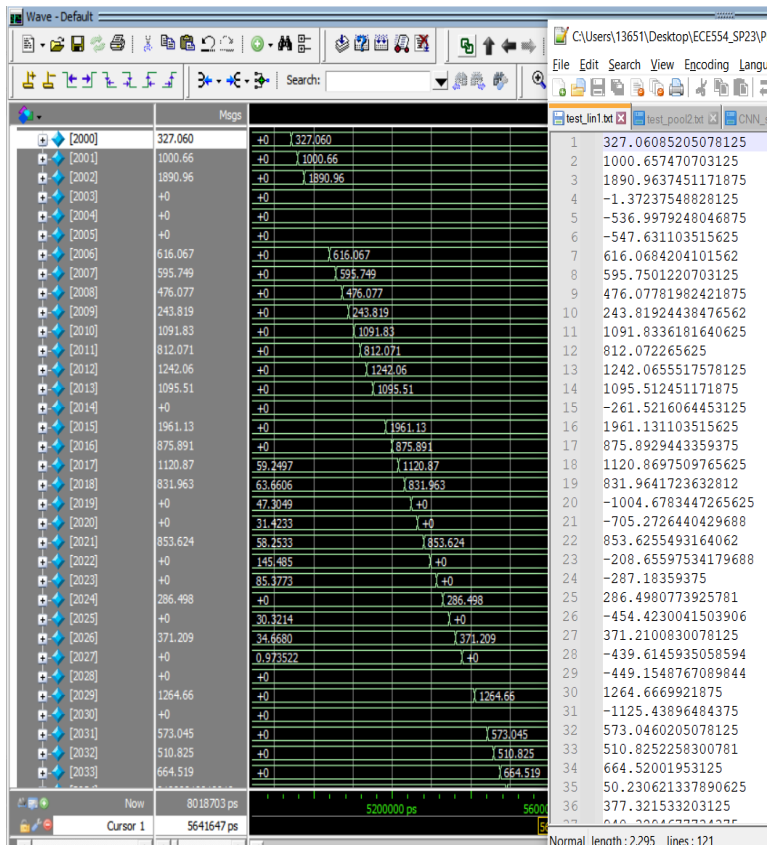

**Figure 12:** More intermediate layer validation.

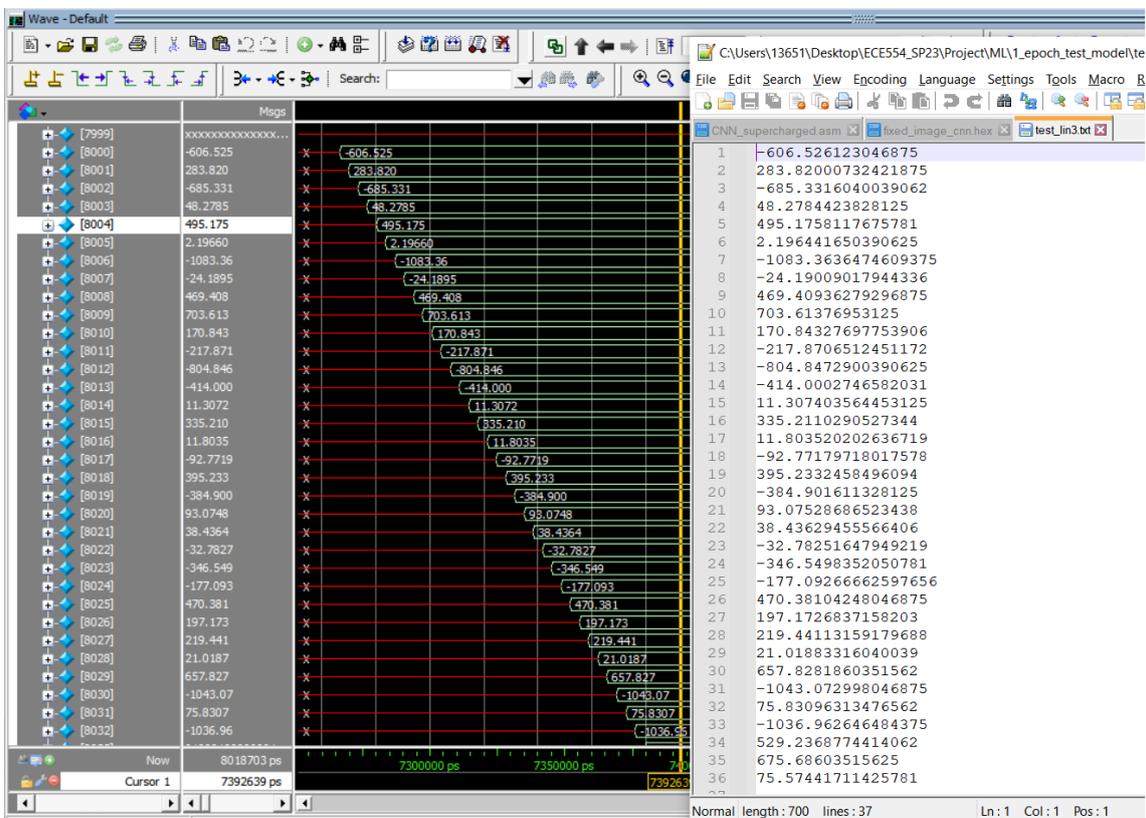

**Figure 13:** Final output validation of the 36 scores.

# 8. Conclusion and Discussions

In conclusion, our handwriting recognition system implemented on the Altera DE1 FPGA Kit demonstrated high accuracy, efficiency, and flexibility with various ML models for handwriting recognition. Our IEEE-754-compliant RISC processor, ISA, custom assembly API, and automated validators could serve as a comprehensive platform for future developments of similar ML applications on FPGAs. In the end, we want to share our reflections on this project's future directions and potential impacts.

## 8.1. New Processor Architecture for ML Acceleration

One approach to further optimize our work involves enhancing the ISA and the processor architecture to better cater to the requirements of ML acceleration. Subsequently, it was discovered that certain pre-designed instructions remained unused in the assembly code and were seldom employed in other ML algorithms. Deliberate attention given to the ISA can substantially enhance the efficiency of the architecture, thereby improving performance. Furthermore, it is important to note that ML algorithms, including the CNN model utilized in our project, entail extensive computations that exhibit parallelizability without any data dependencies. This characteristic renders superscalar architecture particularly advantageous at the architectural level, leading to significant performance gains.

## 8.2. Potential New Features at the Firmware Level

The primary rationale behind our decision to implement a portion of the ML layers in firmware, rather than relying on a dedicated hardware module for performing complete calculations with fixed dimensions, stems from concerns regarding flexibility for future enhancements. By adopting this approach, we maintain the ability to seamlessly transition towards improved FPGA boards or more advanced ML architectures while minimizing the need for extensive modifications to our hardware modules. Instead, modifications to the firmware and simple parameter adjustments within the hardware modules hold the potential for achieving edge detection of individual characters, synchronous recognition of multiple characters, and classifications of more complex alphabets, thereby facilitating higher accuracy, broader capability, and enhanced real-world applications.

## 8.3. The Trend of Edge Computing Using ASIC Devices

The core paradigm behind this design is Edge Computing, which aims to bring computation and data storage closer to where the data is generated. By processing data locally and transmitting only the necessary data, edge devices reduce latency and bandwidth requirements, sometimes eliminating the need for data transmission. This approach improves reliability and scalability by distributing computing resources efficiently without relying on a centralized cloud infrastructure. It also enhances security by keeping critical data within a hardware system closer to its source. Edge computing applications are rapidly emerging in various industries, including smart grid systems, autonomous vehicles, and industrial manufacturing. Our design on the FPGA could eventually be implemented as ASIC devices for commercial use. This would significantly reduce the overheads, relax memory limitations, and improve power consumption.

## 8.4. Floating-Point Arithmetic in FPGA Machine Learning Hardware

Our current implementation is fully based on floating-point arithmetic to achieve these ML algorithms on FPGAs. However, we realized that there could be a potential performance boost if fixed-point arithmetic were utilized. Thus, based on our observations during development, we summarized some advantages and shortcomings of FP architectures in ML compared to fixed-point solutions below.

**Pros:**
(1) **Representable Range:** Given the same number of bits, floating-point numbers allow for representing both infinitesimally small and astronomically large values, allowing for a broader spectrum of numerical magnitudes. This flexibility is particularly useful in ML, where datasets often contain diverse numerical values with varying scales.
(2) **Standardization and Easy Verification:** Floating-point numbers adhere to well-defined standards, which ensures consistency across different platforms and programming languages. This standardization enables the interoperability and portability of ML models and algorithms. Many existing ML software libraries, such as PyTorch and TensorFlow, utilize FP values throughout their implementations [12]. Supporting closer arithmetic, even only 32-bit, can make the verification steps more automatic and efficient.
(3) **Industry Trend:** Floating-point operations are optimized in modern computer architectures, making them faster and more efficient compared to other number representations [13]. This efficiency is crucial for large-scale ML tasks that involve extensive matrix multiplications and gradient computations. Various industry leaders, such as Nvidia, are utilizing different FP hardware for ML-purposed GPU designs [12]. As a project for prototyping purposes, this FPGA-based implementation should support as many functionalities as those high-end tools do when possible, and users could choose not to utilize the FP features at the firmware level.

**Cons:**
(1) **Precision Loss:** Floating-point numbers have limited precision due to the finite number of bits used for representation. As a result, rounding errors and precision loss can accumulate during complex computations, potentially impacting the accuracy of ML models. Careful consideration is necessary when dealing with numerical stability and avoiding issues like catastrophic cancellation.
(2) **Inexact Representations:** Certain decimal numbers cannot be represented exactly in floating-point format, leading to rounding errors. This behavior can be problematic when dealing with applications that require high precision, such as financial calculations or scientific simulations.
(3) **Timing Concerns:** Since floating-point operations involve multiple stages of shorter addition, multiplication, and shifting, the overall critical path could be more significant than a fixed-point chip design for ML. This could cause difficulties in timing closures, especially when mapping to low-power or low-cost ASIC products.

We reckon that for handwriting recognition purposes, converting all kernels, weight matrixes, and inputs from 32-bit floating-point to 32-bit fixed-point representation through proper scaling and data manipulation may not significantly impact the overall accuracy of our CNN architecture. However, under time constraints, we were not able to verify this theory and assess its precession.